# Magneto-convective models of red dwarfs: constraints imposed by the lithium abundance

J. MacDonald and D. J. Mullan

*Dept. of Physics and Astronomy, University of Delaware, Newark DE 19716 USA*

**ABSTRACT**

Magnetic fields impede the onset of convection, thereby altering the thermal structure of a convective envelope in a low mass star: this has an effect on the amount of lithium depletion in a magnetized star. In order to quantify this effect, we have applied a magneto-convective model to two low mass stars for which lithium abundances and precise structural parameters are known: YY Gem and CU Cnc. For both stars, we have obtained models which satisfy empirical constraints on the following parameters: *R, L,* surface magnetic field strength, and Li abundance. In the case of YY Gem, we have obtained a model which satisfies the empirical constraints with an internal magnetic field of several megagauss: such a field strength is within the range of a dynamo where the field energy is in equipartition with rotational energy deep inside the convection zone. However, in the case of CU Cnc, the Li requires an internal magnetic field which is probably too strong for a dynamo origin: we suggest possible alternatives which might account for the reported Li abundance in CU Cnc.

## 1. INTRODUCTION

### 1.1. Magneto-convective modeling of low-mass stars

In a previous paper, MacDonald & Mullan (2014: hereafter MM14) reported on an application of magneto-convective modeling to three M dwarfs. When we apply the adjective "magneto-convective" to our modeling, it has the following specific meaning: the onset of convection in a stellar model is *not* determined by the usual Schwarzschild criterion on the temperature gradient, $d\ln T/d\ln p > (d\ln T/d\ln p)_{ad}$. Instead, in our magneto-convective models, the onset of convection is determined by a physics-based criterion derived by Gough & Tayler (1966: hereafter GT). The GT criterion quantifies the following process: in the presence of a vertical magnetic field, the onset of convection in an electrically conducting medium does *not* occur as easily as it does in a non-magnetic environment. GT showed that magneto-convection sets in only when $d\ln T/d\ln p$ exceeds $(d\ln T/d\ln p)_{ad}$ by a finite amount δ. According to GT, δ is roughly (within factors of order unity) equal to $B^2/8\pi p$, i.e. the ratio between the local magnetic pressure and the local gas pressure.

An essential feature of the GT criterion, with its associated alteration of the temperature *gradient* in the convective zone of a magnetized star, is that the *radial profile of temperature* inside such a star is systematically different from the temperature profile in a non-magnetic star. Specifically, other things being equal, the temperature at a certain depth in a magnetic star model is found to be *lower* than in a non-magnetic star model at the same depth. In the case of a star which has a radiative core, it will in particular be true that conditions at the base of the convection zone are cooler in the magnetic star. But this conclusion is not confined merely to the base: it applies throughout the convection zone. Magnetic fields result in a change in the overall thermal structure of a low-mass star.

The 3 stars which were discussed in detail by MM14 were selected for modeling because Torres (2013) had identified them as eclipsing systems in which the stellar masses and radii have been determined more precisely than for any other M dwarfs. As a result, these stars present the toughest challenges to obtaining credible precise models of low mass stars. (Torres also cited a fourth star GU Boo which has a precisely determined mass and radius, but its age is unknown, and this precludes calculation of meaningful evolutionary models.) MM14 demonstrated that magneto-convective models which provide good fits to the measured radii *R*, luminosities *L*, and surface magnetic field strengths $B_{surf}$ can be obtained in a restricted region of a certain plane which is defined by just two parameters: $\delta$ (as defined by GT) and $B_{ceiling}$. The latter quantity encapsulates the fact that dynamo activity in a star cannot generate magnetic field strengths in excess of some limiting value. The actual value of $B_{ceiling}$ in a "real star" is likely determined by parameters such as rotational speed, convective velocity, diffusivity, etc. Although an exact value of $B_{ceiling}$ inside a star is unknown, we shall offer estimates of upper limits in Section 4 below, in order to test our models for consistency.

MM14 showed that, in the best fitting magneto-convective models of CM Dra, YY Gem, and CU Cnc, the values of $B_{surf}$ in the models are consistent with empirical estimates of $B_{surf}$ based on measurements of the X-ray luminosity $L_X$.

This suggests that the magneto-convective model of MM14 is consistent with (at least) three types of observational data: *R, L,* and $B_{surf}$ (i.e. $L_X$).

**1.2. Lithium abundances: a further constraint on magneto-convective modeling**

In this paper, we explore if the magneto-convective model as applied to M dwarfs is also consistent with one further observational parameter: the abundance of the element lithium, assuming that thermal convection is the only mixing process. As it turns out, Li abundances are available for only 2 of the stars among the 4 prime candidates identified by Torres (2013): YY Gem and CU Cnc.

Barrado y Navascués et al. (1997) measured the equivalent width of the YY Gem Li 6707.8 Å line to be 65 mÅ. For adopted parameters, $T_{eff}$ =3185 K and log *g* = 4.5, they determined from the curve of growth (Soderblom et al. 1993) the Li abundance *A*(Li) = 12+Log[*N*(Li)/*N*(H)] = 0.11 ± 0.43. Since this adopted $T_{eff}$ is significantly lower than that determined by Torres & Ribas (2002; TR02), $T_{eff}$ =3820, we have re-calculated the Li abundance by interpolating in the results of the synthetic spectra calculations by Pavlenko et al. (1995), using the TR02 temperature and equivalent width 65 mÅ. We find *A*(Li) = 0.05 ± 0.15. To allow for other uncertainties such as that associated with the equivalent width measurement and to encompass the error given by Barrado y Navascués et al. (1997), we adopt *A*(Li) = 0.05 ± 0.50. Bonsack (1961) finds Castor Aa has a solar Be/Fe ratio, and so we assume it has a solar system Li abundance, *A*(Li) = 3.28 ± 0.05 (Lodders 2010). Adopting this as the initial Li abundance for YY Gem, its Li depletion is Δ*A*(Li) = -3.23 ± 0.50.

In his high resolution spectra of CU Cnc A and B, Ribas (2003) reported for both components equivalent widths of ~50 mÅ for the Li I line at $\lambda$6708Å, and estimated the Li abundances to be *A*(Li) ≈ −1.1, based on the results of Pavlenko et al. (1995). The adopted temperatures are $T_{eff}$ = 3160 ± 150 for CU Cnc A and $T_{eff}$ = 3125 ± 150 for CU Cnc B**.** Interpolating in the Pavlenko et al. (1995) results, we determine that the uncertainty in the Li abundance arising from the $T_{eff}$ error bars is 0.5 dex. Assuming that CU Cnc is a proper motion companion of Castor (see section 3.1), the Li depletion for each of the CU Cnc components is $\Delta A(\text{Li}) = -4.4 \pm 0.5$.

We use Li depletion, rather than the Li abundance, to compare with our models: the reason is that there exists a small difference between our initial Li abundance (mass fraction $1.00 \times 10^{-8}$) and the solar system Li abundance (mass fraction $1.06 \times 10^{-8}$).

Since the components of YY Gem are essentially identical in mass, we assume that they have the same Li abundance. However, the components of CU Cnc are slightly different in mass (0.43 and 0.39 $M_\odot$), and their structure is sufficiently different to impact the evolution of the surface Li abundances, even in the absence of magnetic field effects. In modelling the depletion of Li, we consider in detail only the more massive component CU Cnc A because it has a shallower surface convection zone than CU Cnc B. The presence of Li in CU Cnc B poses a more challenging constraint on stellar models than its presence in CU Cnc A.

Why is lithium relevant to our models? Because destruction of Li depends on convective transport of material down to deep inner layers of the star where local temperatures are hot enough to allow effective proton capture. Given that magnetic fields alter the thermal structure of a low-mass star in the sense of reducing the local temperature at any given radial location, a star with a magnetic field should cause the Li to be transported down into layers where the local temperatures are cooler than those in a non-magnetic star (other things being equal). As a result, the magnetic star should experience a lesser amount of Li destruction (assuming that there are no non-convective mixing processes). Using the same $\delta$-$B_{ceiling}$ plane as in MM14, we ask: does the empirical Li abundance point towards a restricted region in this plane which overlaps with the regions already discovered by MM14? Or does the Li abundance point to a region in the plane which *does not overlap* with the MM14 regions?

From a historical perspective, the present paper is not the first attempt to explore how magnetic fields in cool stars may alter the Li abundances. In the case of cool evolved stars, MacDonald & Mullan (2003) suggested that the upward buoyancy of magnetic fields should contribute to less severe Li depletion in rotating red giants, although no detailed models were obtained. In pre-main-sequence members of the β Pictoris moving group (BPMG), MacDonald & Mullan (2010) applied the GT criterion to obtain quantitative models of how isochrones and the rate of depletion of $^7$Li abundances are affected by magnetic fields. The magnetic models obtained for BPMG stars were found to have lower luminosity for a given central temperature than in the absence of magnetic fields. As a result, the BPMG stars contract more slowly and $^7$Li is depleted at a lower rate. This has a measurable effect on increasing the "lithium-age" of the stars. Feiden & Chaboyer (2013) applied their magneto-convective model to investigate whether the Li abundance constraints in several M dwarfs can be satisfied by models which fit the observed luminosities and radii. In the GT models which we report here, magneto-convective effects are treated in a different way from that which is used by Feiden & Chaboyer: in particular, we use a different assumption about the radial profile of the magnetic field strength in the stellar interior. Is there any way to verify our assumption of a certain radial profile of $B(r)$ inside a star? In general, the answer is that such a verification would be very difficult to obtain. However, in the case of one particular star (the Sun), the answer is more positive: the profile $B(r)$ is testable, using $p$-mode data. The MDI experiment on SOHO has provided frequencies of multiple $p$-modes in the Sun extending over an entire activity cycle: the data show that there are systematic shifts in frequency of each mode between solar minimum and solar maximum (e.g. Rabello-Soares 2012). Using those data, Mullan et al (2012) have shown that the structural changes in the Sun between solar minimum and solar maximum are consistent with our choice of $B(r)$ inside the Sun. A more recent paper by Somers & Pinsonneault (2014) discusses the extent to which radius inflation by magnetic fields in pre-main-sequence stars can alter the depletion of Li: however, Somers & Pinsonneault do not provide a detailed physical model of magnetic stars. An even more recent paper (still in preprint form as of early November 2014) shows that the presence of dark spots on a star's surface leads to systematically older ages for stars based on the lithium depletion boundary (Jackson & Jeffries 2014).

### 1.3. Plan of the paper

In Sections 2 and 3, we first re-capitulate the results obtained by MM14 for the two stars in their sample (YY Gem and CU Cnc) for which Li abundances are currently available. We present the discussion in the context of the $\delta$-$B_{ceiling}$ plane, showing (A) the locations on that plane which are consistent with empirical $R$ and $L$ (as reported by MM14). Then we overplot (B) the locations (on the same plane) which we have found (in this paper) to be consistent with the empirical Li abundance. The key question is: do the allowed locations (A) and (B) in the $\delta$-$B_{ceiling}$ plane overlap? In Section 4 we examine how magnetic fields and rotation might have varied in the course of evolution. Conclusions are in Section 5.

## 2. YY GEMINORUM

### 2.1. Setting the results in context

In order to prepare for the discussion in this paper, it is important at first to illustrate (in Fig. 1) a tool which was used in MM14 to present the results. The tool is a "two-parameter plane" defined by $\delta$ and $B_{ceiling}$: this plane is useful to demonstrate where, in this parameter space, we have obtained magneto-convective models which are consistent with various empirical quantities. In MM14, the number of empirical quantities we were interested in fitting was at first limited to two: *R* and *L*. (We then tested the models for consistency with a third parameter: $B_{surf}$.) Numerical values of *R* and *L* for each component of YY Gem were taken from TR02. The empirical values are $R = 0.6191 \pm 0.0057$ R$_\odot$ and $L = 0.0701 \pm 0.0095$ L$_\odot$: these are the values we used in MM14, and also in the present paper. As regards the age of YY Gem, TR02 cite an age of $370 \pm 40$ Myr for Castor A and B: since YY Gem is a member of the Castor sextuplet system, we use an age of 370 Myr in our modeling of YY Gem.

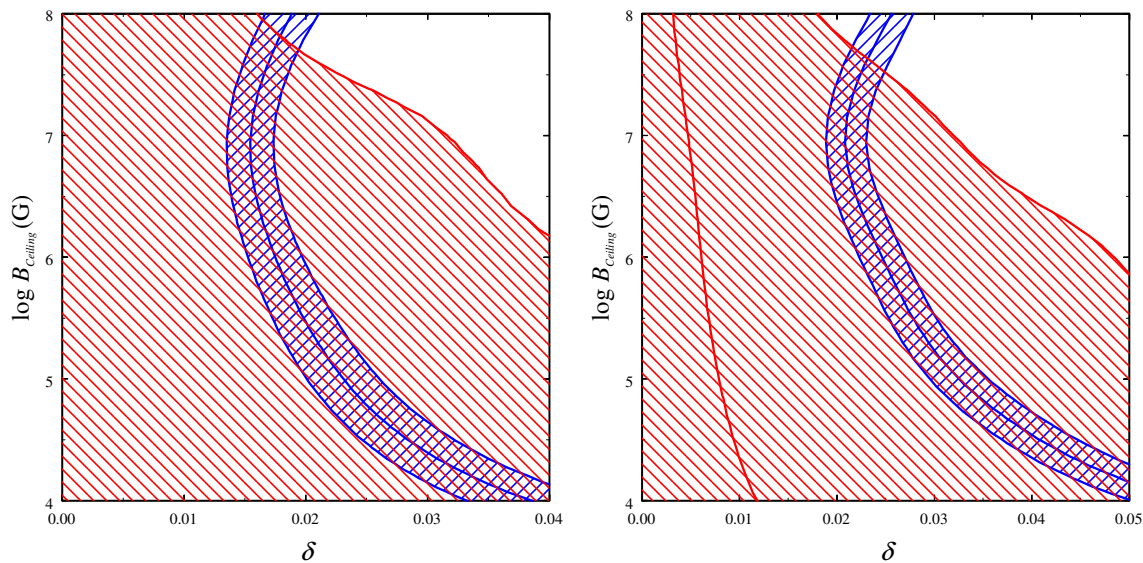

Figure 1. The $\delta$-$B_{ceiling}$ plane for 370-Myr-old magneto-convective models of YY Gem. Models consistent with observed ranges of $R$ ($L$) lie in the blue (red) shaded region of the plane. Cross-hatched (purple) region: models fit $R$ and $L$ values simultaneously. Models are illustrated for mixing length parameter $\alpha = 1.0$ (left-hand panel) and $\alpha = 1.7$ (right-hand panel).

As regards the heavy element abundance of YY Gem, theoretical isochrones (TR02) suggest a value "quite close to solar", [Fe/H] = $0.02 \pm 0.04$, while based on a spectroscopic study, TR02 report [Fe/H] = $0.1 \pm 0.2$ in the photosphere. In the *corona* of YY Gem, X-ray data indicate that the elemental abundances are lower than solar in the quiescent state, but during flares the abundances increase (Guedel et al. 2001): coronal abundances in the Sun are known to differ from the photospheric values by factors of several, depending on the First Ionization Potential (FIP) of the element. These "FIP effects" are presumably due to MHD phenomena operating in partially ionized plasma in the lower atmosphere (e.g. Arge & Mullan 1998). As a result of the FIP effect, it is not a simple matter to relate coronal abundances to photospheric abundances (or interior abundances): thus, the X-ray data do not help us in choosing a value of [Fe/H] which is relevant for evolutionary modeling of the stellar interior. In view of these results, we have used [Fe/H] = 0.0 for YY Gem in what follows.

In order to appreciate the presentation of results in Figure 1, it is important to note one aspect of the empirical quantities: the fractional uncertainty in radius (±0.9%) is more than 10 times smaller than the fractional uncertainty in luminosity (±14%). The reason for the difference is clear. On the one hand, the linear dimensions of an eclipsing binary can be determined with high precision from timing measurements. On the other hand, to determine luminosity, information is required about visual magnitudes, parallaxes, and effective temperatures, all of which are subject to a variety of observational uncertainties.

For completeness, it is worthwhile to mention what the "fractional uncertainties" mentioned above correspond to in terms of the usual "standard deviation" $\sigma$ which emerges from a discussion of formal internal errors in the analysis. In TR02 (their Section 4.1.2), there is an extensive discussion of this topic, based on fits of the standard Wilson-Devinney (WD) algorithm (including dark spots) to YY Gem eclipse data: "The uncertainties given in our table were not adopted from the (often underestimated) formal probable errors provided by the WD code but instead from numerical simulations and other considerations. Several sets of starting parameters were tried in order to explore the full extent of the parameter space. In addition, the WD iterations were not stopped after a solution was found, but the program was kept running to test the stability of the solution and the geometry of the $\chi^2$ function near the minimum. The scatter in the resulting parameters from numerous additional solutions yielded estimated uncertainties that we consider to be more realistic and are generally *several times larger* than the internal errors." That is, estimates of formal errors may not be realistic in the case of stars such as YY Gem where systematic effects (e.g. starspots) may contribute significantly, In a discussion of systematic errors (TR02: their Section 5.1), it is noted that "the uncertainties given above for our mass determinations (0.8%) are strictly internal errors. Even though we have made every effort to minimize systematics in our radial velocity measurements, residual errors affecting the *accuracy* are bound to remain at some level." TR02 cite another estimate of masses in YY Gem made by other investigators, and find differences of order $4\sigma$ in terms of the internal errors. A third investigation leads to a mass estimate which differs by 1.7% from the results of TR02: TR02 refer to this 1.7% discrepancy as "a $2\sigma$ difference". Thus, when Torres (2013) cites 4 systems as having mass determinations which are reliable to 3% or better, the uncertainties should be considered as corresponding to not merely formal internal errors, but as also including (to some reasonable extent) certain systematic errors, especially those which arise in the presence of starspots. Their final word on this topic is enlightening: "The message we wish to convey to the reader with the arguments presented in this section is the difficulty in reaching *accuracies* (as opposed to a *precision*) in the absolute masses much better than 1% for a system with the characteristics of YY Gem. It is entirely possible that some of the astrophysical effects mentioned above, such as spottedness, may ultimately set the limit for stars of this class."

In Figure 1, we replicate the $\delta$-$B_{ceiling}$ plane for YY Gem, including the results obtained by MM14. The notation is as follows. In the left-hand panel, we show results obtained with mixing length parameter $\alpha$ = 1.0. In the right-hand panel, we show results for models in which we used $\alpha$ = 1.7 (the typical value which is needed for our non-magnetic model to fit the present Sun.) Inside each panel, the color scheme is as follows. Magneto-convective models which are consistent with the empirical *radius R* all lie within the area which is shaded in *blue*. Magneto-convective models which are consistent with the empirical *luminosity L* all lie within the area shaded in *red*.

Why are the blue and the red areas in Fig. 1 so different in shape and area? Because of the large differences in the precisions with which $R$ and $L$ have been measured. The $R$ value is so tightly constrained (0.9%) that only a narrow (blue) area of the $\delta$-$B_{ceiling}$ plane is consistent with the data. On the other hand, the less stringent limits on $L$ (14%) have the effect that results consistent with the $L$ data can be found over a much broader (red) area of the $\delta$-$B_{ceiling}$ plane.

The cross-hatched purple area in Figure 1 denotes pairs of values of $\delta$ and $B_{ceiling}$ for which MM14 found models which are simultaneously consistent with *both R and L.* We note the following two important aspects of the results in Fig. 1.

(a) The empirical values of *both R and L* in YY Gem can be fitted simultaneously only if $B_{ceiling}$ is less than 100 MG (where the blue lines cross over the right-most red line, for $\alpha$ = 1.0). For $\alpha$ = 1.7, $B_{ceiling}$ cannot exceed 30-50 MG. Also, acceptable simultaneous fits exist only for $\delta$ values $\geq$ 0.014-0.02: these minimum values for $\delta$ occur at $B_{ceiling}$ = 4-8 MG.

(b) Despite the broad uncertainty attached to the "best" numerical value of *one* of our two parameters (namely, $B_{ceiling}$: uncertain by possibly as much as $10^4$ ), MM14 found that as regards our determination of numerical values for our *second parameter* ($\delta$), the limits on the permissible range were much smaller. Empirical constraints are such that we are simply *not* permitted to have $\delta$ values which are arbitrarily large. To see this, note that the numerical value of $\delta$ $\left( \approx B^2/8\pi p \right)$ in any best-fit magneto-convective model immediately leads, in combination with $p$ at the stellar surface (also provided by the model), to a surface magnetic field strength $B_{surf}$.) The value of $B_{surf}$ in turn leads to a predicted coronal X-ray flux which can be compared with observational data. MM14 showed that the empirical X-ray fluxes from the stars CM Dra, YY Gem, and CU Cnc are in fact consistent with the $\delta$ values obtained by our best-fit magneto-convective models, provided that the $\delta$ values were not in excess of 0.022, 0.04, and 0.046 for CM Dra, YY Gem, and CU Cnc respectively. Therefore, in our analysis of Fig. 1, we do not need to examine values of $\delta$ in excess of 0.04 for YY Gem: such values of $\delta$ would lead to such strong $B_{surf}$ values that they would lead to X-ray fluxes in disagreement with the observed values. As a result, we do not need to consider values of $\delta$ beyond the right-hand side which is plotted in Fig. 1. Neither do we need to consider $B_{ceiling}$ values below 10 kG: even an old, slowly rotating, weakly active stars such as our Sun is capable of generating fields in its interior as strong as 10-15 kG (Browning 2008), so we expect the rapid rotators CM Dra, YY Gem, and CU Cnc to have fields of at least those magnitudes (or probably considerably larger) in the interior.

As a result, the limits plotted in Fig. 1 are adequate to cover the relevant (and permitted) areas of parameter space for our purposes.

MM14 found that successful models were all found to have values of $\delta$ which lay within a narrow range: the uncertainty in the "best" value of $\delta$ was found to be no more than a factor of 2. The most striking aspect of the results in MM14 was precisely the following feature: the models allow us to evaluate one parameter ($\delta$) fairly reliably *even when the second parameter ($B_{ceiling}$) is subject to large (several orders of magnitude) uncertainties.* And since $\delta$ can be constrained by means of an observable (the X-ray flux), whereas $B_{ceiling}$ cannot (yet) be, this can be considered as a positive feature of our magneto-convective modeling approach.

**2.2. Models which address the lithium abundance**

In figure 2, using the models we have computed using the GT criterion, we illustrate the predicted degree of Li depletion in the surface layers of a YY Gem component as a function of age. The observed Li abundance (relative to solar) is illustrated by means of horizontal lines in the two panels in Fig. 2. In non-magnetic models (plotted in red in Fig. 2), the Li abundance is reduced to below the observed amount at an age of only 11 – 18 Myr, depending on the mixing length ratio. But as we have already mentioned (Section 1.2 above), in models where magnetic fields are present, the temperature at each radial location in the convection zone is reduced relative to the value in the non-magnetic model. These reduced temperatures have the effect that depletion of Li by proton capture does not proceed as far in magnetic models as in the non-magnetic model. In the limiting case, the process of Li depletion comes to a complete halt at an age when the temperatures throughout the convection zone become too low for proton capture to occur: this age can be seen in Fig 2 as a flattening out of the Li abundance towards a constant value. For our best fitting magneto-convective models, Li depletion ends before age 70 Myr and the uncertainties in the age of YY Gem have no effect on the predicted Li abundance. According to the results plotted in Fig. 2, agreement between our magneto-convective models and the observed Li abundance is obtained for a narrow range of $\delta$ values (= 0.015-0.025). It is encouraging that this range of $\delta$ values is consistent with the upper limit on $\delta$ (=0.04) which is imposed on YY Gem by observed X-ray fluxes (see Section 2.1(b) above).

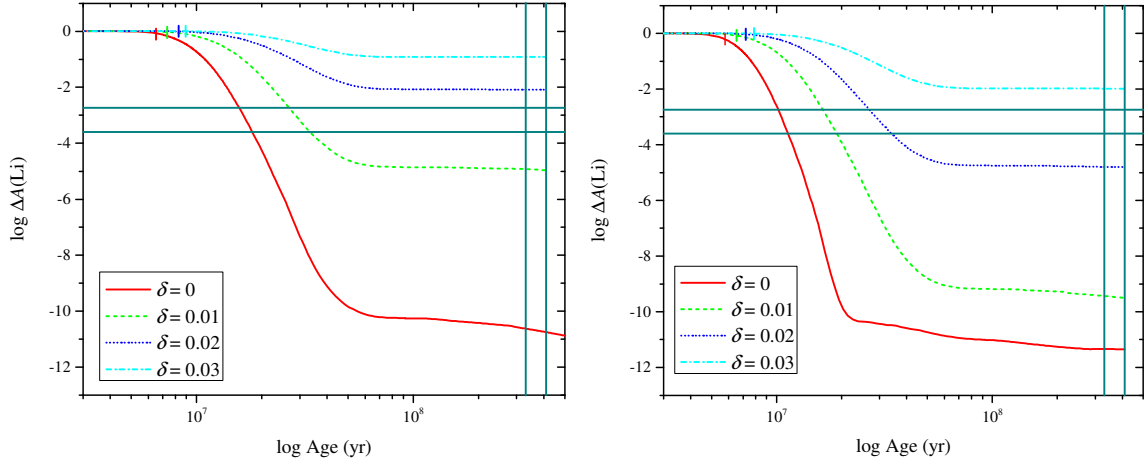

Figure 2. Evolution of the Li depletion at the surface of magneto-convective models of a YY Gem component for $\alpha = 1.0$ (left) and $\alpha = 1.7$ (right). The ceiling value of the magnetic field is 1 MG. The horizontal lines show the limits on the Li depletion from spectroscopy. The long vertical lines near the right-hand border show the range of ages determined from isochrone fitting to the companion stars Castor Aa and Ba. The short vertical lines on each track show the age at which a radiative core develops.

In Fig. 3, we present results, using the same 2-parameter plane as in Fig. 1, for our magneto-convective models of YY Gem which now include constraints from the Li abundance. Results in Fig. 3 are plotted for a snapshot at 370 Myr, and assuming that the numerical value of $\delta$ remains constant throughout the evolution. In order to highlight the differences between MM14 models and the current Li models, the magneto-convective models which are consistent with the observed Li abundance lie in a region of the $\delta$-$B_{ceiling}$ plane which is plotted in *green*.

Remarkably, Fig. 3 indicates that there exists an area in the $\delta$-$B_{ceiling}$ plane where the green region (Li) overlaps with the purple region ($R, L$) which had been previously identified in Fig. 1. This indicates that our magneto-convective models can simultaneously satisfy *three* constraints on the parameters of YY Gem: *R, L,* and Li abundance. Note that the blue (radius) constraint and the green (Li) constraint in Fig. 3 overlap well: this overlap may arise because these constraints may not be completely independent, since Li depletion depends on internal temperatures, and the latter scale as *M/R*.

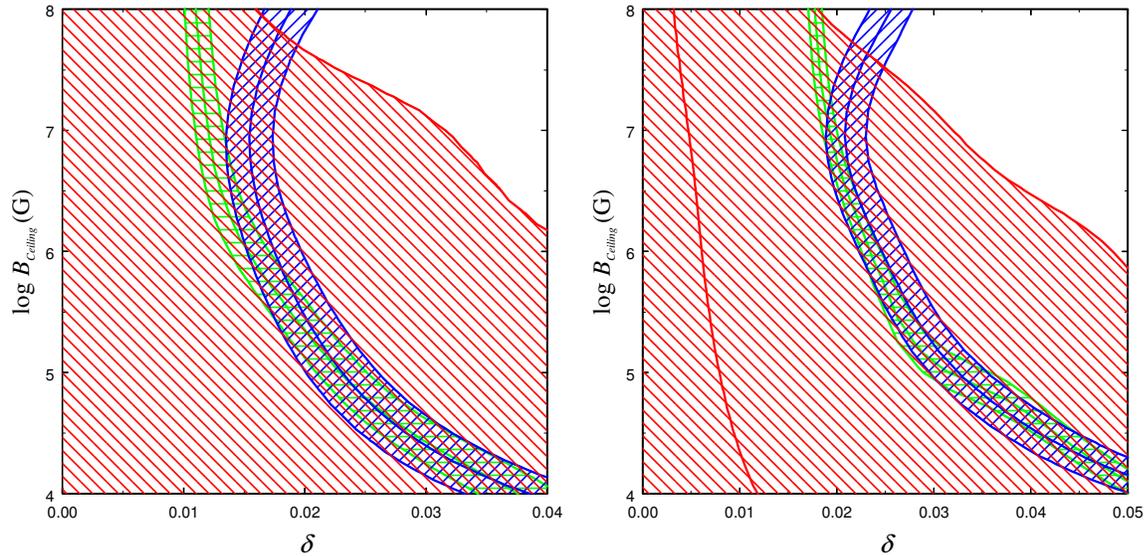

Figure 3. The $\delta$-$B_{ceiling}$ plane for YY Gem. Filled contours of the Li depletion (green), radius (blue) and luminosity (red) at age 370 Myr for $\alpha = 1.0$ (left) and $\alpha = 1.7$ (right). Acceptable simultaneous fits to 3 observational constraints correspond to where all three shaded regions overlap.

The region of triple overlap in Fig. 3 sets an upper limit on $B_{ceiling}$ of 10 MG in YY Gem. MM14 have already shown that such values of $B_{ceiling}$ are well within the reach of an internal dynamo in YY Gem. And the minimum value of $\delta$ in the region of triple overlap is 0.014 (if $\alpha = 1.0$) and 0.02 (if $\alpha = 1.7$). These limits on $\delta$ are identical to those obtained in MM14. Therefore, when combined with the surface gas pressure in the models, the limits on $\delta$ in Fig. 3 are still entirely consistent with the $B_{surf}$ values reported in MM14.

Comparison of the two panels in Figure 3 reveals a significant aspect of Li depletion in our models. In Fig. 3 (left), the width of the Li-constraint (green) area in the $\delta$-$B_{ceiling}$ plane spans a range of 0.002-0.003 in $\delta$ at the top of the figure. In contrast, the green area in Fig. 3 (right) is narrower, with a width of no more than 0.001-0.002 in $\delta$ at the top of the figure. This difference in width can be associated with the difference in the steepness of the curves in Figure 2 (left) and Figure 2 (right). In Fig. 2 (left), the red curve falls off rather gradually, with a maximum slope of -17 in the log-log plot. But in Fig. 2 (right), the red curve falls off more steeply, with a maximum slope is -23. Thus, the internal structure of the models in Fig. 2 (left) are such that once the central temperature of the collapsing pre-main sequence star becomes high enough for the onset of Li depletion, the internal conditions are such that the depletion proceeds at a relatively slow pace, extending over tens of millions of years. The slowness of this pace maps into a relatively broad area in the $\delta$-$B_{ceiling}$ plane (Fig. 3 left). In contrast, the steeper curves of Li depletion in Fig. 2 (right), where Li depletion occurs over a shorter interval of time (roughly 10 million years), map into a narrower region in the $\delta$-$B_{ceiling}$ plane (Fig. 3 right). We will return to this aspect of our models when we discuss CU Cnc below.

## 3. CU CANCRI

### 3.1. Model results

Ribas (2003) has suggested that CU Cnc belongs to the Castor moving group. Unfortunately, Mamajek et al. (2013) have argued that the Castor moving group (including such widely separated putative members as Fomalhaut and Vega) may not exist as a physically connected system of co-eval stars moving through space together. However, Mamajek (2014) has suggested (see Section 3.2 below) that an interesting similarity does exist between the space motions of the particular system CU Cnc and Castor itself. This does not necessarily mean that CU Cnc and Castor are strictly co-eval. But it is at least suggestive, especially since the systems are not too far apart on the sky. Accordingly, in order to compute models of CU Cnc, we adopt an age for the system which is comparable to that of Castor. Recognizing that we have no strong reason to use precisely the same age as Castor (370 Myr), we round this number up by roughly the uncertainty (quoted by TR02), and use the round number of 400 Myr for the age of CU Cnc. Moreover, in view of the link with Castor, we assume [Fe/H] = 0.0 for CU Cnc, as we assumed for YY Gem.

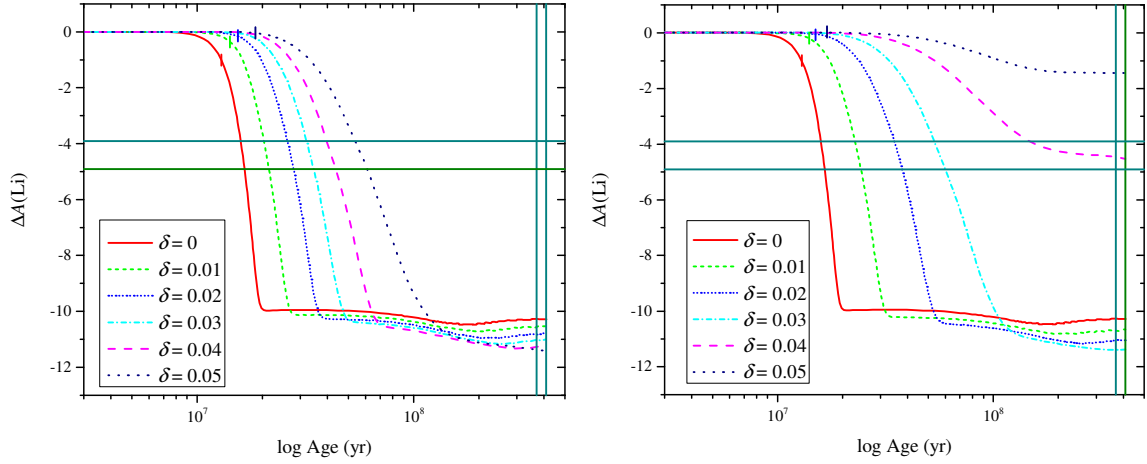

Figure 4. Evolution of the Li depletion at the surface of magneto-convective models of CU Cnc A for $\alpha = 1.7$. The ceiling value of the magnetic field is 1 MG (left panel) and 100 MG (right panel). The horizontal lines show the limits on the Li depletion from spectroscopy. The long vertical lines show the range of ages determined from isochrone fitting to the proper motion companion stars Castor Aa and Ba. The short vertical lines show the age at which a radiative core develops.

In figure 4, the predicted degree of Li depletion in the surface layers of CU Cnc A is shown as a function of age for magneto-convective models with $\alpha = 1.7$. Results shown in the left panel are for models in which $B_{ceiling}$ = 1 MG. In the right panel $B_{ceiling}$ = 100 MG. For the non-magnetic models (red lines), lithium is found to become depleted below the observed amount at an age of only 17 Myr. In the $B_{ceiling}$ = 1 MG models with $\delta$ in the range required by the radius constraint, lithium is depleted before age 60 Myr. Agreement with the observed lithium abundance is obtained for $B_{ceiling}$ = 100 MG and a narrow range of $\delta$ around $\delta \approx 0.04$ (consistent with the upper limit of 0.046 reported by MM14 see Section 2.1(b) above).

In Figure 5, we present results analogous to those in Fig. 3 except that now, the models and data refer to CU Cnc A, and the ordinate axis extends to larger values (300 MG) than was the case for YY Gem). Notice the same topology of the red and blue regions in the plot: a narrower region in blue (reflecting more precise empirical estimates of *R*), and a broader region in red (reflecting less precise estimates of *L*). For $\alpha = 1.0$ and also for $\alpha = 1.7$, the empirical *R* (blue) and *L* (red) do not allow us to set any significant limits on $B_{ceiling}$ within the range which is plotted. According to Fig. 5, $B_{ceiling}$ could be in excess of 100 MG, or less than 100 kG. We will have to rely on other arguments, independent of *R* and *L* if we wish to set limits on $B_{ceiling}$ in CU Cnc A.

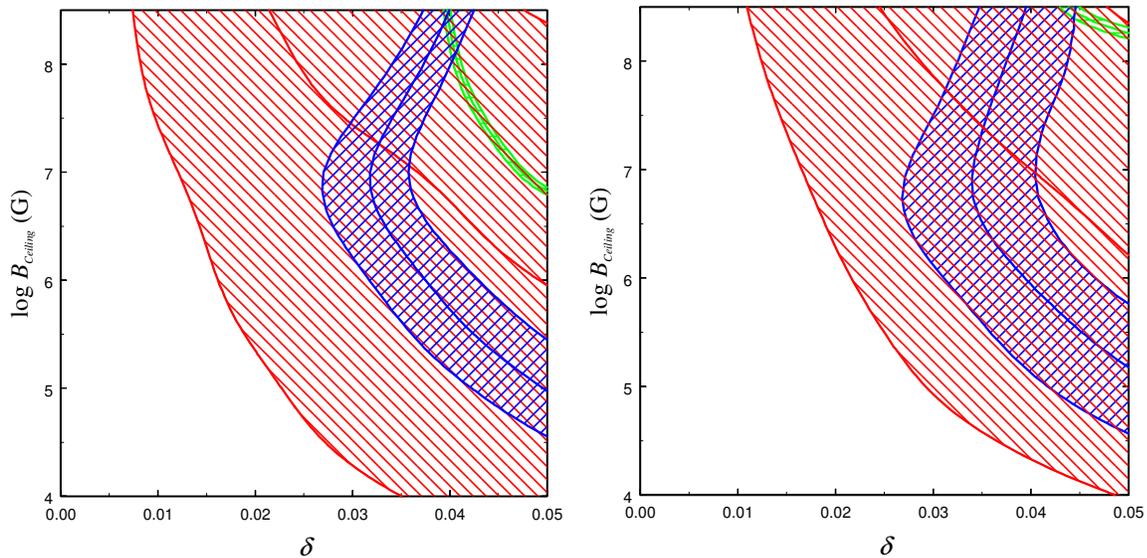

Figure 5. The $\delta$-$B_{ceiling}$ plane for magneto-convective models of CU Cnc A at age 400 Myr. The empirical ranges of *R* (*L*) correspond to the blue (red) shaded regions of the plane. Models have $\alpha = 1.0$ (left) and $\alpha = 1.7$ (right). Models which are consistent with the empirical Li abundance are the narrow bands plotted in green near the upper right hand corner.

To illustrate constraints imposed by the empirical Li abundance in CU Cnc A, we once again in Fig. 5 use green to plot the region in the plane where magneto-convective models are consistent with the Li data. The green regions are not as easy to spot for CU Cnc A (in Fig. 5) as they were for YY Gem (in Fig. 3). The green regions for CU Cnc A are found to be very restricted, and confined to the upper right-hand corner of each panel in Fig. 5.

Why are the green regions in Fig. 5 so narrow? Following on from the discussion of the results for YY Gem above, we associate this narrowness in CU Cnc with the increased steepness of the Li depletion curves in Fig. 4. In Fig. 4 (left) the maximum slope of the red curve is of order -50 in the log-log plot. Thus, the process of Li depletion (once it gets going) in a star with a mass of 0.43 solar masses (as in Fig. 4) occurs in a much shorter time-scale than in a star with a mass of 0.6 solar masses (as in Fig. 2). And in Fig. 4 (right), the steepness is even more extreme: the maximum slope of the red curve is of order -70. Such steep slopes have the effect that even a slight change in local conditions in a stellar model can lead to a large difference in the amount of Li depletion. As a result, the curves in Fig. 4 map into very small areas in the $\delta$-$B_{ceiling}$ plane (Fig. 5).

Nevertheless, the results in Fig. 5 indicate that, remarkably, there does in fact exist a (small) region of triple overlap in the two-parameter space between the green (Li) results and the purple (*R, L*) results.

However, in the case of CU Cnc A, the values of $B_{ceiling}$ which are permitted by our magneto-convective models, turn out to be very large. For $\alpha = 1.0$, the Li abundance, *R* and *L* measurements require $B_{ceiling} > 100$ MG. For $\alpha = 1.7$, $B_{ceiling} > 300$ MG is needed. It is difficult to see how dynamo operation could produce such strong fields. In the event that a dynamo cannot produce such large fields, we might be led to conclude that, in the case of CU Cnc A, our magneto-convective model is unable simultaneously to satisfy the constraints of Li abundance, *R, L,* and $B_{surf}$ for CU Cnc A.

Interestingly, in the context of an independent calculation of stellar convection in the presence of a magnetic field, Feiden & Chaboyer (2013) have also reported difficulties in reconciling the field strengths required by their model of CU Cnc with empirical data: their model fields are also required to be too strong to be consistent with the data. This result is consistent with our finding that internal fields in CU Cnc are required to be very strong. (Feiden & Chaboyer also report similar excessive field strengths for YY Gem, although in the present work [see Fig. 3] we have not found such excesses with YY Gem.)

In modelling CU Cnc A, we have assumed [Fe/H] = 0. The heavy element abundance of Castor determined by TR02 is [Fe/H] = 0.1 ± 0.2, and so we have considered models with a range of heavy element abundances. Due to increased opacity, higher [Fe/H] gives larger radii and reduces the discrepancy between the observed and model radii. However the radius of our $\alpha = 1.0$, [Fe/H] = 0.3 non-magnetic model is still smaller than the observed radius. Also the problem remains with surface Li abundance being depleted during the pre-main sequence phase at age 17 Myr. Again, the observed radius can be matched by including magnetic field effects. For example, a good fit is obtained by our $\alpha = 1.0$, [Fe/H] = 0.3 magnetic model with $B_{ceiling}$ = 1 MG and $\delta$ = 0.01. However, as is the case with the solar composition models, this ceiling value of the magnetic field strength is too low to prevent Li being destroyed during the pre-main sequence phase. In fact, we find for $\alpha = 1.0$, [Fe/H] = 0.3 magnetic models that satisfy the radius constraint, Li is depleted during the pre-main sequence phase even if no ceiling is applied to the magnetic field strength. Lower values of [Fe/H] give smaller model radii and the magnetic field parameter must be increased to match the observed radius. However, the differences between [Fe/H] = -0.1 models and [Fe/H] = 0.0 models are small. For example, our $\alpha = 1.0$, [Fe/H] = -0.1 magnetic models with $B_{ceiling}$ = 1 MG match the observed radius for $\delta$ = 0.0318 – 0.0427, whereas if [Fe/H] = 0.0 the required range is $\delta$ = 0.0315 – 0.0424. The Simultaneous matching of the radius and Li abundance constraints for [Fe/H] = -0.1 is possible only if $B_{ceiling}$ > 60 MG. Although this limit is lower than that found from our [Fe/H] = 0.0 models, it still is higher than field strengths expected from a dynamo.

**3.2. The lithium discrepancy in magneto-convective models of CU Cnc**

We can suggest a number of possible resolutions to the incompatibility between our magneto-convective models and the empirical Li abundance in CU Cnc:

(1) Perhaps the reported Li abundance determination is incorrect. If in fact Li is completely depleted, we find that surface fields as low as 420 G are sufficient to give the required radius inflation. Such fields would fit perfectly into the range of observed surface fields reported by Morin et al. (2010) for stars in the mass range 0.2 - 0.4 $M_\odot$ (such as CU Cnc B: and CU Cnc A almost falls into this range).

(2) Mamajek et al (2013) have argued that the Castor moving group might not in fact exist as a well-defined group of stars. If this is the case, then it might seem that we could not depend on the age and heavy element abundance of Castor to determine the age and heavy element abundance of CU Cnc. (Note: uncertainty of age does not apply to YY Gem, which is a physical member of the Castor system itself, where age is well-determined from evolutionary fits to the more luminous members.) On the other hand, Mamajek (2014) has recently informed us that the space velocity of the CU Cnc system itself is actually quite close to that of Castor: moreover, CU Cnc is not far apart from Castor in the sky, no more than 18 degrees. On this basis, Mamajek suggests that even if the identity of a "Castor moving group" may no longer be defensible, the individual star CU Cnc moves in space so similarly to Castor that one might be permitted to consider CU Cnc as physically related to Castor. To the extent that this suggestion is valid, an age of order 400 Myr and near solar heavy element abundances could be acceptable for CU Cnc after all.

If CU Cnc is not associated with Castor, then the age and composition of CU Cnc are not well constrained. For our non-magnetic solar composition models of CU Cnc A, the observed degree of Li depletion for a star with a mass of 0.43 $M_\odot$ is obtained at ages of no more than 13 – 17 Myr. At such a young age, the radius is about 0.75 – 0.8 $R_\odot$, which is significantly *larger* than the observed radius. Reducing the heavy element abundance does not make a major difference. For example, if [Fe/H] = -1, the radius is 0.6 $R_\odot$ at the lithium age. It seems inescapable that based on the Li abundance the age of CU Cnc A cannot be as young as 13 – 17 Myr. The observed radii and luminosities suggest that the components are on, or very close to, the main sequence. In such a case, very large interior fields are needed to suppress Li depletion.

(3) Perhaps the Li on the surface of the components of CU Cnc is in the process of being accreted from a circum-binary disk. In the absence of Li depletion, this would require a mean accretion rate of $2 \times 10^{-14}$ $M_\odot$ yr$^{-1}$. Simulations which include accretion of solar system composition material at a uniform rate onto our model with $\delta = 0.04$ and $B_{ceiling} = 1$ MG model show that the accretion rate must be between $2 \times 10^{-13}$ $M_\odot$ yr$^{-1}$ and $2 \times 10^{-12}$ $M_\odot$ yr$^{-1}$ in order to balance the destruction of Li by proton capture and maintain the Li abundance at the observed level. If accretion is occurring from a gaseous circumstellar disk, the disk has to have had an initial mass greater than about $7 \times 10^{-5}$ $M_\odot$ = $2 \times 10^{-4}$ $M_*$. Alternatively the Li could be supplied by accretion by each component of a single object. For an object that is mainly rocky material, its mass needs to be approximately that of the planet Mercury.

## 4. EVOLUTION OF MAGNETIC FIELDS AND ROTATIONAL PERIODS

So far, we have assumed that the ceiling magnetic field, $B_{ceiling}$, and the magnetic inhibition parameter $\delta$ remain constant throughout the evolution of the star. The radius and luminosity of our main sequence models are mainly determined by the instantaneous values of the magnetic parameters and are insensitive to the prior evolution of these parameters. However, the same is not true for the Li abundance because depletion of Li occurs during the pre-main sequence phase at ages of some tens of millions of years. In this section, we explore whether our assumptions are plausible. We begin by considering the field evolution assuming that the stars rotation period is constant and equal to the currently observed orbital period. Next, we consider how the stellar rotation period might have evolved due to tidal effects.

In the absence of a detailed dynamo model, precise estimates of magnetic field strengths $B$ inside stars are unavailable. But we can set an upper limit on $B$ in a star with an interface dynamo by making the following specific assumption: an upper limit on $B$ is set by enforcing equipartition between the magnetic energy density and the rotational kinetic energy density deep inside the convective envelope. That is, $B_{equi}^2/8\pi = \rho_b V_b^2(rot)/2$, where subscript $b$ denotes values at the base of the convective zone. Is there any way to determine the plausibility of this assumption? We can offer the following argument, based on what is known about the Sun. At the base of the solar convection zone, i.e. at radial location $r \approx 0.7\ R_\odot$, the rotational speed is roughly 0.7 times the surface speed, i.e. $V_b(rot) \approx 1.4$ km/sec. And the local gas density is $\rho_b = 0.19$ g cm$^{-3}$ (e.g. Bahcall et al 2006). These values lead to an upper limit $B_{equi} \approx 220$ kG. Now, Choudhuri & Gilman (1987) have shown that in order to explain why sunspots remain at relatively low latitudes (rather than lying close to the poles), the buoyancy forces on rising magnetic flux tubes must exceed the Coriolis forces: to ensure this, the magnetic field strength at the base of the solar convection must have a lower limit $B_c$. Based on the empirical distribution of sunspots, Choudhuri & Gilman conclude that $B_c \approx 100$ kG. Our estimate of the upper limit on rotationally-induced fields in the sun, $B_{equi} \approx 220$ kG, is consistent with the empirical lower limit $B_c \approx 100$ kG.

In Fig. 6 we show the values we have obtained for $B_{equi}$ as a function of age in our evolutionary magneto-convective models of two M dwarfs. (The stars are assumed to be rigid rotators.) The left panel is for a YY Gem component ($P_{orb} = 0.814$ days), and the right panel is for CU Cnc A ($P_{orb} = 2.77$ days). (Note that both stars, with masses of 0.60 and 0.43 $M_\odot$ respectively, are sufficiently massive that an interface *does* exist in both of our target stars.)

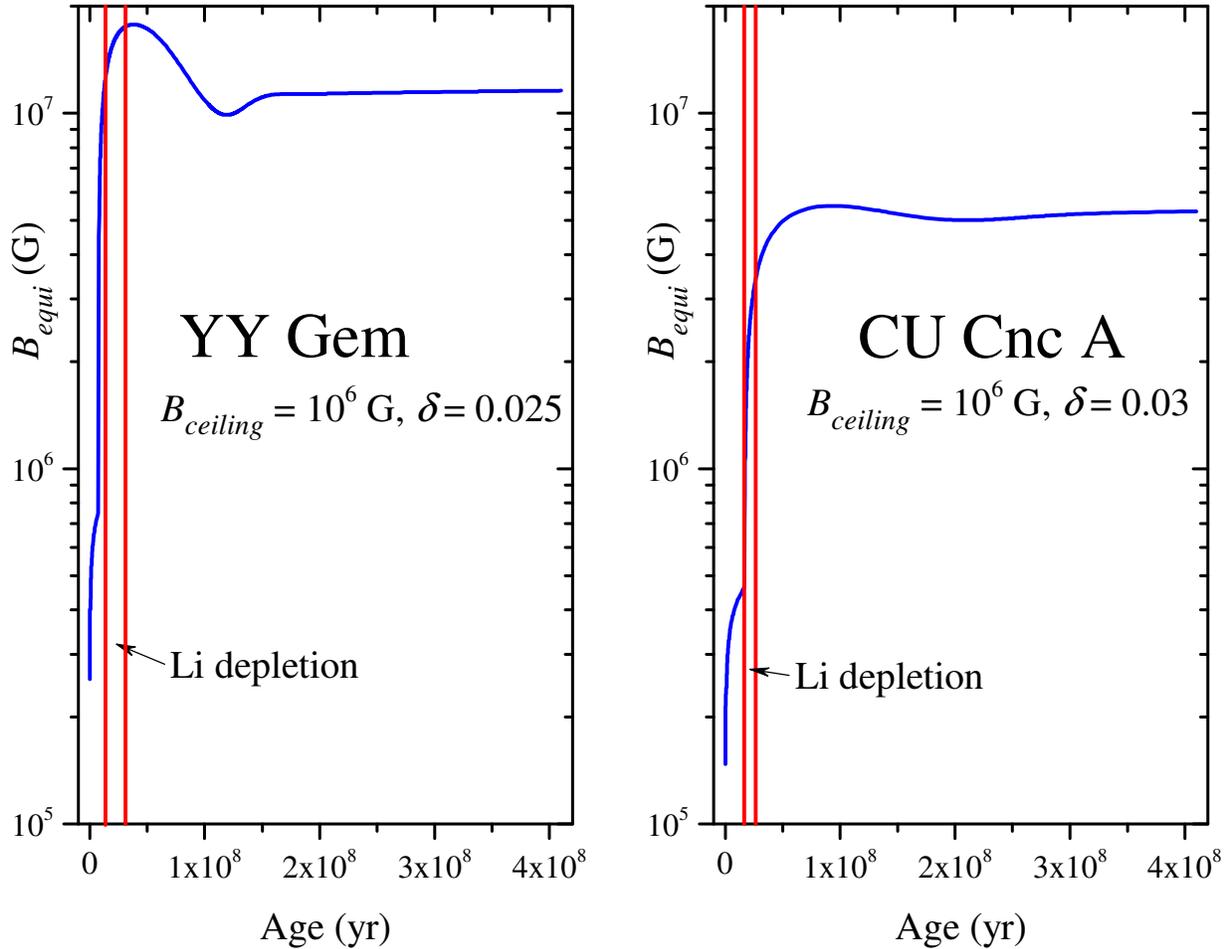

Figure 6. Equipartition magnetic field strength $B_{equi}$ at the base of the outer convection zone as a function of stellar age for our two M dwarf target stars. Local parameters are taken from one particular magneto-convective model which satisfies, in each star separately, the constraints which have been set empirically on $R$ and $L$ for that star.

We have found (see Fig. 3 above) that our magneto-convective models of YY Gem with $B_{ceiling}$ in the range of several MG *can* satisfy empirical constraints on $R$, $L$, $B_{surf}$, and Li. Now we see, in Fig. 6, that such values of $B_{ceiling}$ are not inconsistent with the upper limit indicated by $B_{equi}$. This is an encouraging result for YY Gem. However, for CU Cnc A, Fig. 6 suggests that even with the optimal assumptions which go into calculating $B_{equi}$, we cannot account for the very strong fields (100 MG or more: see Fig. 5 above) which would be required if magneto-convective models are to fit the empirical Li abundance.

So far, in calculating the $B_{equi}$ values in Fig. 6, we have assumed the orbital periods of the YY Gem and CU Cnc binaries remain unchanged as a function of age. Now let us consider how we might relax that assumption. To estimate how much the stellar rotation periods are likely to change during the early evolution of the binary systems, there are two processes which must be included: (i) tides between the two components, and (ii) mass loss from each individual component of the system. In order to model (i), we have solved the equations of Zahn (1989) for the evolution of the stellar angular velocities, orbital semi-major axis and eccentricity due to tidal interactions between the components of a binary. In order to model (ii), we have also included the angular momentum lost in stellar winds using the prescription in Chaboyer,

Demarque & Pinsonneault (1995): the parameter $N$ that appears in their equations (3) and (4) is taken to be $N = 3/2$ to make the angular momentum loss rate independent of the stellar mass loss rate. The stellar properties, including rotational inertia, are given by our stellar evolution calculations. In figure 7, we show the predicted evolution of the orbital period and the axial rotational periods of each star in the binary, assuming initial conditions of a nearly perfectly circular orbit and zero axial rotation rate for each star. The starting age for the period evolution is chosen so that the stars just fit inside their Roche lobes. The initial orbital period is chosen to give the currently observed orbital periods at an age of 370 Myr.

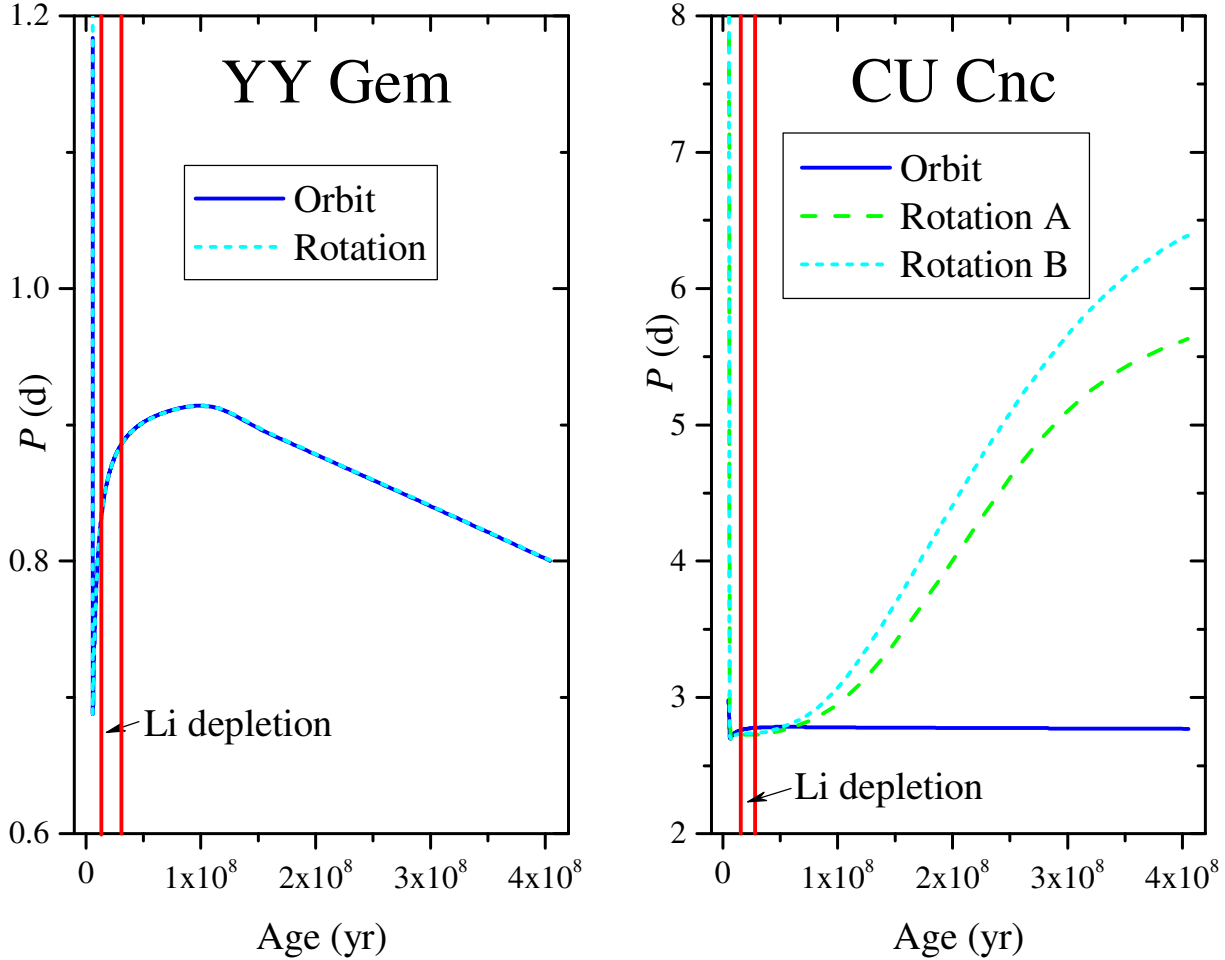

Figure 7. Two target binary systems: evolution of orbital period (in days) and axial rotation periods (in days) on each component as a function of time using the theory of Zahn (1989). In YY Gem, the orbit curve and the rotation curve are indistinguishable except in the upper left-hand corner. But in CU Cnc, the wider separation of the orbit allows significant non-synchronicity to develop.

In the case of YY Gem (currently with an orbital period of 0.8 days), we find that the axial rotation becomes locked to the orbit in the first $10^5$ yr, and remains locked at least to the early main sequence phase. The variation in orbital period from the time span of onset of Li depletion to reaching the main sequence is of order 10%. TR02 have measured the rotational velocities of the YY Gem components, $V\sin i$ = 36.4 and 37.8 km s$^{-1}$ for A and B respectively, which correspond to rotational periods that are 6% and 2% longer than the orbital period respectively. TR02 do not give error bars for their projected rotational velocity measurements, but for typical uncertainties of a few km s$^{-1}$, the rotational periods are consistent with synchronous rotation. By combining the results shown in figures 6 and 7, we conclude that the magnetic field strengths in the YY Gem components might not vary significantly over the time interval relevant to connecting the constraint from the Li abundance to the radius and luminosity constraints at the current epoch. This suggests that our choice of constant values of the parameters $\delta$ and $B_{ceiling}$ throughout evolutionary time in YY Gem should not give rise to significant errors in our results. Also, magnetic ceiling values of up to 10 MG are plausible throughout this whole time interval. We have explored other initial conditions, including eccentric orbits, and find that, because the synchronization time scale is so short, the stellar rotation rates at age 370 Myr are not significantly different. We have also considered a scenario in which the binary was formed when the components have age 100 Myr, at which point the stellar radii are only 10% larger than at age 370 Myr. The circularization and synchronization time scales are 20 Myr and 35 Myr. Even in this case the orbit is circular and the stellar rotation is synchronized with the orbit at age 370 Myr.

However, in the case of CU Cnc, the situation is different. Because of its longer orbital period (currently 2.77 days, i.e. 3 – 4 times longer than the orbital period of YY Gem), we find tidal locking is not maintained: as a result, by the time the stars settle on the main sequence, due to the stellar wind angular momentum loss, their rotation periods are a factor 2 – 2.5 longer than the orbital period. This result is insensitive to the initial conditions used. We do not know of any measurements of the projected rotational velocities for the CU Cnc components and cannot (yet) confirm or deny our conclusion of asynchronous rotation. For CU Cnc, we find that (see Figure 6) the largest plausible $B_{ceiling}$ value is of order 5 MG at age near 100 Myr, with a decrease by a factor 2 at later times. Slower rotations lead to smaller values of $B_{equi}$, no more than 5 MG: such values are too small to be consistent with the values required for $B_{ceiling}$ (>100 MG) if our magneto-convective models are to replicate the empirical Li abundance (see Figs. 4 and 5).

## 5. CONCLUSIONS

We have obtained magneto-convective models for M dwarf stars in two binary systems where the empirical masses and radii are among the most precisely known current values (Torres 2013).

In the case of YY Gem, we have found that magneto-convective models can be found which simultaneously satisfy empirical constraints on *R, L,* and Li abundance. We have previously discovered (in MM14) that we could find consistent solutions between *two* of these constraints (*R* and *L*): and those solutions were also found to be consistent with indirect estimates of surface magnetic field strength $B_{surf}$. So our models of YY Gem might now be considered to satisfy quadruple constraints. (Because of complex interconnections between conditions inside the star, we cannot yet prove that the four constraints are completely independent of one another: we postpone a study of this topic to a future paper.)

As a further consistency check, we have shown in this paper that, in the case of YY Gem, the successful "quadruple constraint" models require internal magnetic fields which *can* be supplied by dynamo action, at least in the context of a dynamo where rotational energy at the base of the convective envelope is effectively converted to magnetic energy. These results are encouraging indications that our magneto-convective modeling is consistent with multiple empirical properties of YY Gem.

In the case of CU Cnc A, it appears at first sight that we have also achieved a measure of success: we have obtained magneto-convective solutions which can in principle simultaneously satisfy the four constraints of *R, L,* surface field strengths, and Li. However, in CU Cnc A, the "successful" models require internal fields which seem to too strong to be generated by reasonable dynamo parameters. In that sense, our model of CU Cnc A is not yet as satisfactory as our model for YY Gem. In Section 3.2 above, we have suggested some possible solutions to this problem.

We do not claim that, because we have fitted Li in one particular system (YY Gem) with our magneto-convective models, the Li problem has been solved in magnetic low-mass stars. There are still free parameters in our modeling: e.g. at any given stellar age, the amount of Li depletion is sensitive to assumptions about mixing length and chemical composition. Thus, better confidence in our magneto-convective models awaits the accumulation of reliable Li data in a larger sample of cool stars with precisely determined masses and radii, and above all, with reliable determinations of the age. Only if the age is known with some confidence can we hope to test evolutionary models in a physically meaningful manner.

**ACKNOWLEDGMENTS**

DJM acknowledges partial support from Delaware Space Grant. We thank the anonymous referee for valuable comments.